\documentclass[conference]{IEEEtran}
\usepackage{graphicx, array, amssymb, psfrag, graphics, setspace, stfloats, psfrag}
\usepackage{amsfonts}
\usepackage[caption=false,font=scriptsize]{subfig}
\usepackage[toc,abbreviations,nonumberlist,nogroupskip,nopostdot]{glossaries-extra}
\glssetcategoryattribute{general}{glossdesc}{firstuc}
\glssetcategoryattribute{abbreviation}{glossdesc}{title}
\usepackage{mathtools}
\usepackage{xurl}

\DeclareMathOperator*{\argmin}{arg\,min}

\newcommand{\m}[1]{\mathbf{#1}}
\newcommand{\E}[1]{\mathbb{E} \left[{#1}\right]}
\newcommand{\Prob}[1]{\text{Pr} \left({#1}\right)}

\newcommand{\myabs}[1]{\left\lvert#1\right\rvert}

\def\scalingFig{0.32}

\newglossary*{nomenclature}{Nomenclature}

\newabbreviation{ISI}{ISI}					{inter-symbol interference}
\newabbreviation{DAC}{DAC}					{digital-{}to-analog converter}
\newabbreviation{ADC}{ADC}					{analog-{}to-digital converter}
\newabbreviation{DSP}{DSP}					{digital signal processing}
\newabbreviation{TX}{TX}					{transmitter}
\newabbreviation{RX}{RX}					{receiver}
\newabbreviation{PSK}{PSK}					{phase shift keying}
\newabbreviation{QAM}{QAM}					{quadrature-amplitude modulation}
\newabbreviation{FEC}{FEC}					{forward error correction}
\newabbreviation{SOP}{SOP}					{State-{}of-polarization}
\newabbreviation{FF}{FF}					{feed-forward}
\newabbreviation{FFEQ}{FFEQ}				{feed-forward equalizer}
\newabbreviation{BER}{BER}					{bit error rate}
\newabbreviation{SNR}{SNR}					{signal-{}to-noise ratio}
\newabbreviation{RSNR}{RSNR}				{required SNR}
\newabbreviation{SNDR}{SNDR}				{signal-{}to-noise-{}and-distortion ratio}
\newabbreviation{SFDR}{SFDR}				{spurious free dynamic range}
\newabbreviation{RC}{RC}					{raised cosine}
\newabbreviation{RRC}{RRC}					{root raised cosine}
\newabbreviation{ENOB}{ENOB}				{Effective number of bits}
\newabbreviation{GD}{GD}					{group delay}
\newabbreviation{CMD}{CMD}					{chromatic dispersion}
\newabbreviation{PMD}{PMD}					{polarization mode dispersion}
\newabbreviation{PDL}{PDL}					{polarization dependent loss}
\newabbreviation{ASE}{ASE}					{amplified spontaneous emission}
\newabbreviation{LMS}{LMS}					{least mean squares}
\newabbreviation{APA}{APA}					{affine projection algorithm}
\newabbreviation{NLMS}{NLMS}				{normalized LMS}
\newabbreviation{MMSE}{MMSE}				{minimum mean square error}
\newabbreviation{CMA}{CMA}					{Constant modulus algorithm}
\newabbreviation{RLS}{RLS}					{recursive least squares}
\newabbreviation{LS}{LS}					{least squares}
\newabbreviation{LO}{LO}					{local-oscillator}
\newabbreviation{CR}{CR}					{carrier-recovery}
\newabbreviation{ASIC}{ASIC}				{application-specific integrated circuits}
\newabbreviation{FIR}{FIR}					{finite impulse response}
\newabbreviation{IIR}{IIR}					{infinite impulse response}
\newabbreviation{DD-LMS}{DD-LMS}			{decision-directed least mean squares}
\newabbreviation{DD}{DD}					{decision-directed}
\newabbreviation{CS-DAC}{CS-DAC}			{current-steering DAC}
\newabbreviation{LSB}{LSB}					{least-significant bit}
\newabbreviation{MSB}{MSB}					{most-significant bit}
\newabbreviation{DNL}{DNL}					{Differential non-linearity}
\newabbreviation{INL}{INL}					{Integral non-linearity}
\newabbreviation{DGD}{DGD}					{differential group delay}
\newabbreviation{FFT}{FFT}					{fast-Fourier transform}
\newabbreviation{IFFT}{IFFT}				{inverse fast-Fourier transform}
\newabbreviation{DFT}{DFT}					{discrete Fourier transform}
\newabbreviation{IDFT}{IDFT}				{inverse discrete Fourier transform}
\newabbreviation{FT}{FT}					{Fourier transform}
\newabbreviation{MSE}{MSE}					{mean square error}
\newabbreviation{HD}{HD}					{hard decision}
\newabbreviation{SD}{SD}					{soft decision}
\newabbreviation{LDPC}{LDPC}				{low-density parity check}
\newabbreviation{CW}{CW}					{continuous wave}
\newabbreviation{PBC}{PBC}					{polarization beam combiner}
\newabbreviation{MIMO}{MIMO}				{multiple-input {}and multiple-output}
\newabbreviation{SISO}{SISO}				{single-input {}and single-output}
\newabbreviation{OPGW}{OPGW}				{optical ground wire}
\newabbreviation{ZF}{ZF}					{zero-forcing}
\newabbreviation{CAZAC}{CAZAC}				{Constant amplitude zero auto-correlation}
\newabbreviation{CFO}{CFO}					{carrier frequency offset}
\newabbreviation{MA}{MA}					{moving average}
\newabbreviation{DE}{DE}					{differential evolution}
\newabbreviation{SA}{SA}					{simulated annealing}
\newabbreviation{DEM}{DEM}					{Dynamic element matching}
\newabbreviation{LUT}{LUT}					{lookup table}
\newabbreviation{DP}{DP}					{dynamic programming}
\newabbreviation{DPC}{DPC}					{digital pre-compensation}
\newabbreviation{NN}{NN}					{neural network}
\newabbreviation{MLSE}{MLSE}				{maximum likelihood sequence estimation}
\newabbreviation{LE}{LE}					{linear equalizer}
\newabbreviation{DFE}{DFE}					{Decision–feedback equalizer}
\newabbreviation{THP}{THP}					{Tomlinson-Harashima precoding}
\newabbreviation{HW}{HW}					{hardware}
\newabbreviation{PS}{PS}					{pilot sequence}
\newabbreviation{SW-LS}{SW-LS}				{sliding window least squares}
\newabbreviation{RD-Kalman}{RD-Kalman}      {radius-directed Kalman}		
\newabbreviation{TCM}{TCM}					{trellis coded modulation}
\newabbreviation{CER}{CER}					{constellation expansion ratio}
\newabbreviation{SER}{SER}					{symbol error rate}
\newabbreviation{AWGN}{AWGN}			    {additive white Gaussian noise}
\newabbreviation{A-RLS}{A-RLS}              {approximated recursive least squares}
\newabbreviation{SQNR}{SQNR}                {signal-{}to-quantization-noise ratio}
\newabbreviation{TS}{TS}                    {Training Sequence}
\newabbreviation{Pol}{Pol}                  {Polarization}
\newabbreviation{RMS}{RMS}                  {root mean square}
\newabbreviation{PMF}{PMF}                  {probability mass function}
\newabbreviation{CMF}{CMF}                  {cumulative mass function}
\newabbreviation{PDF}{PDF}                  {probability distribution function}
\newabbreviation{GCS}{GCS}                  {geometric constellation shaping}
\newabbreviation{PCS}{PCS}                  {probabilistic constellation shaping}
\newabbreviation{DM}{DM}                    {distribution matching}
\newabbreviation{SpSh}{SpSh}                {sphere shaping}
\newabbreviation{CCDM}{CCDM}                {constant composition distribution matching}
\newabbreviation{SM}{SM}                    {shell mapping}
\newabbreviation{ESS}{ESS}                  {enumerative sphere shaping}
\newabbreviation{PCDM}{PCDM}                {prefix-free code DM}
\newabbreviation{HiDM}{HiDM}                {hierarchical DM}
\newabbreviation{MPDM}{MPDM}                {multiset-partition distribution matching}
\newabbreviation{CDF}{CDF}                  {cumulative distribution function}
\newabbreviation{KL}{KL}                    {Kullback-Leibler}
\newabbreviation{AC}{AC}                    {arithmetic coding}
\newabbreviation{IID}{IID}                  {independent and identically distributed}

\graphicspath{{Fig/}} 

\begin{document}

\title{Architecture Design for Rise/Fall Asymmetry Glitch Minimization in Current-Steering DACs}

\author{
    \IEEEauthorblockN{Ramin Babaee\IEEEauthorrefmark{1}, 
    Shahab Oveis Gharan\IEEEauthorrefmark{2}, 
    and Martin Bouchard\IEEEauthorrefmark{1}} 
    \IEEEauthorblockA{\IEEEauthorrefmark{1}School of EECS, University of Ottawa, Ottawa, ON  K1N 6N5, Canada, \{ramin.babaee,bouchm\}@uottawa.ca}
    \IEEEauthorblockA{\IEEEauthorrefmark{2}Ciena Corp., Ottawa, ON K2K 0L1, Canada, soveisgh@ciena.com}
}

  

\maketitle

\begin{abstract}
Current-steering \gls*{DAC} is a prominent architecture that is commonly used in high-speed applications such as optical communications. One of the shortcomings of this architecture is the output glitches that are input dependent and degrade the dynamic performance of the DAC. We investigate \gls*{DAC} glitches that arise from asymmetry in the fall/rise response of DAC switches. We formulate a glitch metric that defines the overall DAC performance, which is then used to find a novel DAC weighting scheme. Numerical simulations show that the proposed architecture can potentially provide a significant performance advantage compared to the segmented structure.     
\end{abstract}


\section{Introduction}

Current-steering DACs are widely used in high-speed applications \cite{toumazou_1993, 9863991}. The non-linear distortions caused by circuit element mismatches degrade the overall performance. Glitches are one of the several distortions that play a significant factor in designing efficient DACs. Glitches are transients at the output of the \gls*{DAC} which occur when the input signal changes from one value to another. The error caused by a glitch is a non-linear function of the input encoding \cite{RTSC}. The main causes of glitches are timing mismatch among current switches, asymmetry in the rise/fall behavior, and charge feed-through and capacitive coupling \cite{RTSC}. 

A binary \gls*{DAC} architecture offers a low complexity circuit and efficient area but suffers from significant distortions. In contrast, thermometer-coded design greatly improves the performance, but at the cost of significant complexity. A segmented architecture proposes a hybrid design that consists of binary weighting for \gls*{LSB}s and unary weighting for \gls*{MSB}s \cite{986168, 628767}. The trade-off between complexity and performance offers flexibility for the \gls*{DAC} circuit design. The segmented architecture offers an improved glitch performance, compared to binary weighting. 


\gls*{DEM} techniques exploit the redundancy in unary architectures to convert the distortion into white noise. This is achieved by random selection of current elements in the circuit. Many \gls*{DEM} techniques have been proposed in the literature \cite{RTSC,5420027,913021,9485113}. These approaches significantly improve \gls*{SFDR}. However, \gls*{SNDR} remains unchanged. Several other approaches are also studied in the literature that target glitches. Return-to-zero schemes are an efficient way of suppressing glitches \cite{1696302,315200}. However, they are not suitable for many high-speed applications. The authors in \cite{9025028} propose using dithering and low-pass filtering for converting short high-amplitude glitches to long-duration low-amplitude transients. In \cite{9438957}, a method for compensating switching glitches is presented, which involves generating a complementary amount of glitches at the output of the DAC to offset the original glitch. Additionally, \cite{7360216} introduces a dynamic capacitance compensation technique for reducing glitches in binary DACs.

In our paper \cite{AE_paper}, we introduced the concept of weighting optimization for minimizing statistical amplitude errors. In this paper, we focus on the glitches caused by different rise/fall settling behavior. We propose a weighting optimization that minimizes the glitch power at the receiver. We present several representation selection algorithms with different computational complexities. We then compare the efficiency of the proposed architecture with the traditional segmented weighting scheme through simulations. 

\textbf{Notation:} Vectors and scalars are represented by bold letters and non-bold italic letters, respectively. $\m{B}_i$ represents the $i$-th element of vector $\m{B}$. $(\cdot)^\text{T}$ indicates transpose of a vector/matrix. Symbol $\E{\cdot}$ is used for statistical expectation operation.

\section{Current-Steering DAC}
Let's consider an ideal $N$-bit current-steering \gls*{DAC} which is composed of $L$ current sources, where the weight of $i$-th source is denoted as $\m{B}_i$. The signal at the output of \gls*{DAC} at sampling index $n$ can be represented as
\begin{equation}
x[n] = \m{W}^\text{T}(x[n]) \m{B},
\end{equation}
where $\m{W}(x)$ is a binary vector of size $L$ which represents the input $x$, and $\m{B}$ is a basis vector of the same size containing the weights of all current sources. A current source of weight $\m{B}_i$ is usually implemented by $\m{B}_i$ parallel unit current sources. A binary \gls*{DAC} has $N$ switches where $\m{B}_i = 2^i I_u$ and $I_u$ is the output current of a single-unit current source. In a unary architecture, the number of switches is $2^N-1$ and $\m{B}_i =1$ for all sources.

\section{Basis Design}

\begin{figure}[!tb]
\centering
\footnotesize
\psfrag{Switching ON transient}{Switching on transient}
\psfrag{Switching OFF transient}{Switching off transient}
\psfrag{time}{Time [s]}
\psfrag{T}{$T$}
\psfrag{Output current}{\hspace{-5ex} Normalized output current} 
\includegraphics[scale=\scalingFig]{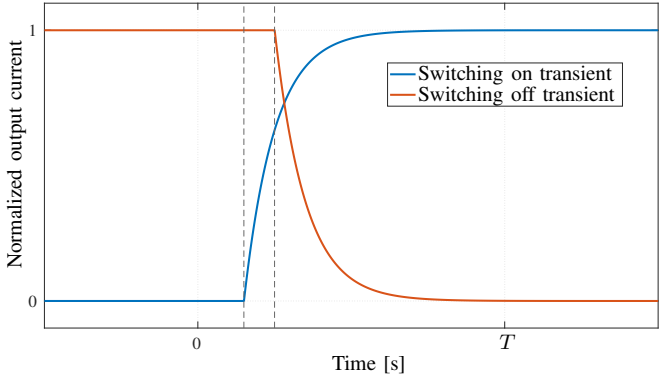}
\caption{Settling behavior during switching on and off.}
\label{fig:ag_s_on_off}
\end{figure}

Glitches occur when the \gls*{DAC} input changes from one value to another. Following \cite{OnOffGlitchModeling}, we assume all the switches of the \gls*{DAC} have the same settling transient, except that the on and off switching waveforms are skewed. An example is illustrated in Fig. \ref{fig:ag_s_on_off} which shows a static time offset between on and off transients. The authors in \cite{OnOffGlitchModeling} derived a metric for the glitch error caused by a transition from input $x$ to input $y$. The glitch error is given as
\begin{equation}
e_g(x,y) = K \myabs{\m{W}^\text{T}(y) - \m{W}^\text{T}(x)} \m{B},
\label{eq:ag_err}
\end{equation}
where $K$ is a constant, $\myabs{\cdot}$ denotes the element-wise absolute value of a vector, and $x=\m{W}^\text{T}(x)\m{B}$ and $y=\m{W}^\text{T}(y)\m{B}$. For notation simplicity, we drop constant $K$ for the rest of this paper.

For a complete basis, e.g. binary weighting, there is only representation for each codeword. However, an over-complete basis such as segmented \gls*{DAC} offers redundancy and therefore, a codeword may be represented by multiple binary vectors. Let $\mathcal{R}(x)$ denote the set of all possible representations for input $x$, i.e.,
\begin{equation}
\mathcal{R}(x) = \left\{\m{W} | \m{W}^\text{T} \m{B} = x\right\}.
\end{equation}
For a complete basis, the glitch metric defined as the expected value of glitch error power is given as
\begin{equation}
\begin{aligned}
\mathbb{E}_{x,y} \big[|e_g(x,y)|^2\big] & = \sum_{x=0}^{2^N-1} \sum_{y=0}^{2^N-1} \Prob{x,y} \\ & 
\hspace{10ex} \Big(\myabs{\m{W}^\text{T}(y) - \m{W}^\text{T}(x)} \m{B}\Big)^2,
\label{eq:ag_metric}
\end{aligned}
\end{equation}
where $\Prob{x,y}$ is the transition probability from $x$ to $y$. In the case of an over-complete basis, since there might be multiple representations for a codeword, the metric could be modified to use the best representation of $y$ that minimizes Eq. (\ref{eq:ag_err}). Thus, we have
\begin{equation}
\begin{aligned}
\mathbb{E}_{x,y} \big[|e_g(x,y)|^2\big] &= \sum_{x=0}^{2^N-1} \sum_{y=0}^{2^N-1} \Prob{x,y} \sum_{\m{W}(x) \in \mathcal{R}(x)} \Prob{\m{W}(x)} \\ 
& \min_{\m{W}(y) \in \mathcal{R}(y)} \Big(\myabs{\m{W}^\text{T}(y) - \m{W}^\text{T}(x)} \m{B}\Big)^2,
\end{aligned}
\end{equation}
where $\Prob{\m{W}(x)}$ is the probability distribution of representations of $x$.
The goal is to find an over-complete basis that achieves similar glitch performance as segmented \gls*{DAC} but with fewer elements. We can express the over-complete basis optimization problem as
\begin{equation}
\begin{aligned}
\m{B}_\text{opt} &= \argmin_{\m{B}} \sum_{x=0}^{2^N-1} \sum_{y=0}^{2^N-1} \Prob{x,y} \sum_{\m{W}(x) \in \mathcal{R}(x)} \Prob{\m{W}(x)} \\ 
& \hspace{5ex} \min_{\m{W}(y) \in \mathcal{R}(y)} \Big(\myabs{\m{W}^\text{T}(y) - \m{W}^\text{T}(x)} \m{B}\Big)^2.
\label{eq:ag_op}
\end{aligned}
\end{equation}
For each representation $\m{W}(x) \in \mathcal{R}(x)$, we find the best representation of $y$ that minimizes the glitch error power. Note that the optimization expression (\ref{eq:ag_op}) is a non-linear, non-convex, discrete problem. The dimension of search space is $L$ and each element may have an integer value in the range $[1 \; 2^N]$. Therefore, the maximum size of the search space is $2^{NL}$, which grows exponentially with both $N$ and $L$. Therefore, an exhaustive search is not feasible except for small values of $L$ and $N$, and a numerical heuristic method should be used.

\begin{figure}[!tb]
\centering
\footnotesize
\includegraphics[scale=0.35]{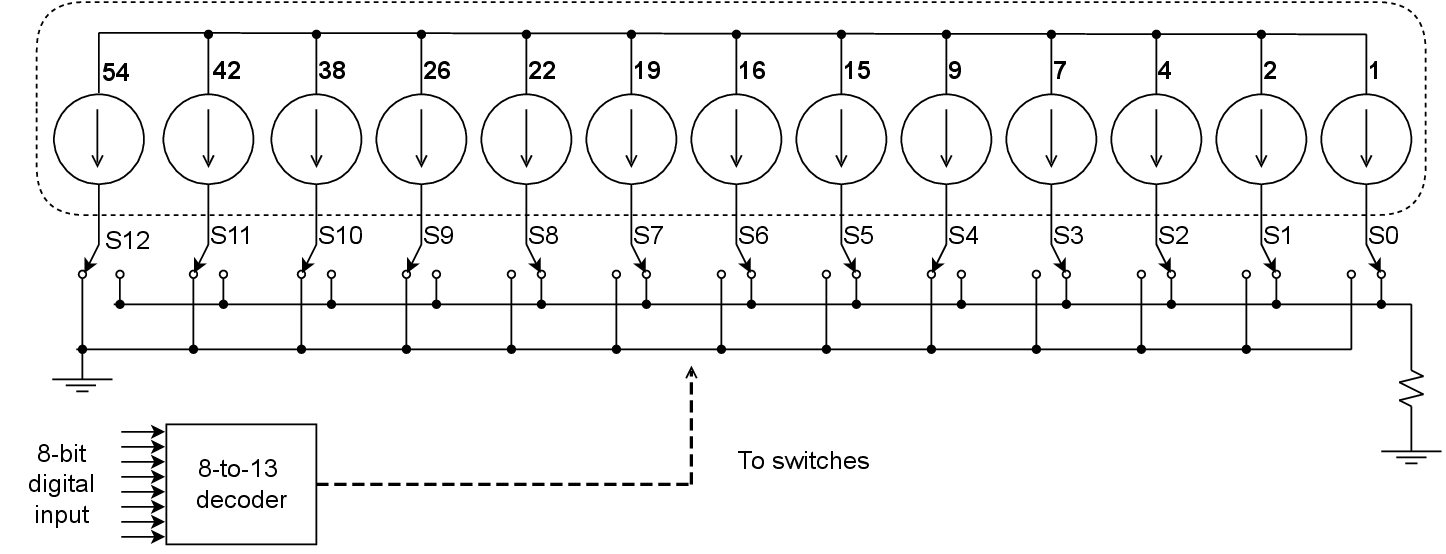}
\caption{Block diagram of the optimized $13$ switches architecture.}
\label{fig:ag_13_architecture}
\end{figure}

Using \gls*{SA} algorithm \cite[Chapter 7]{optimization_book}, the optimized basis for an $8$-bit \gls*{DAC} for a few values of $L$ is computed and is presented in Table \ref{tab:ag_basis}. The optimization algorithm is run 100 times and the best basis is selected from the results. The block diagram of the optimized $13$ switches architecture is illustrated in Fig. \ref{fig:ag_13_architecture}. 
\begin{table}[ht]
\footnotesize
\centering
\begin{tabular}{||l||*{13}{p{0.3em}}||}
\hline
\textbf{Basis Length} & \multicolumn{13}{l||}{\textbf{Optimized Basis}} \\ 
\hline
9  &  1&  2&  4&  8&  16&  31&  43&  69&  81&    &    &    &	\\
10 &  1&  2&  4&  8&  16&  21&  31&  39&  62&  71&    &    &	\\
11 &  1&  2&  4&  8&  13&  18&  26&  30&  38&  54&  61&    &    \\
12 &  1&  2&  4&  8&  11&  16&  20&  25&  27&  35&  48&  58&    \\
13 &  1&  2&  4&  7&   9&  15&  16&  19&  22&  26&  38&  42&  54\\
\hline
\end{tabular}
\vspace{1ex}
\caption{Asymmetric glitch performance optimized basis vectors for an $8$-bit DAC.}
\label{tab:ag_basis}
\end{table}

\section{Representation Selection}

\begin{figure}[!tb]
\centering
\footnotesize
\includegraphics[scale=0.5]{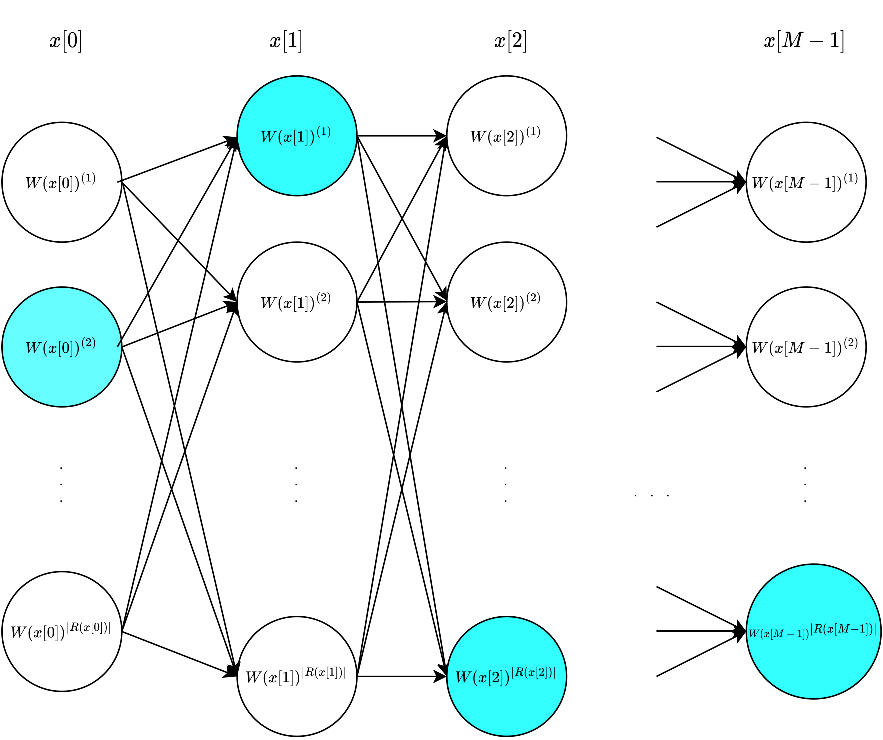}
\caption{Trellis diagram for the Viterbi algorithm. The representations shown in blue represent the optimal path found by the Viterbi algorithm.}
\label{fig:ag_trellis}
\end{figure}

Given an optimized basis vector $\m{B}$ and an input sequence $x[0]$, $x[1]$, ..., $x[M-1]$, the goal is to find the best representations that minimize the glitch error power, i.e., 
\begin{equation}
\argmin_{\m{W}(x[0]),...,\m{W}(x[M-1])} \sum_{m=1}^{M-1} \Big( \big| \m{W}^\text{T}\big(x[m]\big) - \m{W}^\text{T}\big(x[m-1]\big) \big| \m{B}\big)^2.    
\end{equation}

\textit{1) Viterbi (optimal) mapping}: Similar to the theory of hidden Markov models, a \gls*{DP} algorithm such as the Viterbi can be used to find the optimal representations for a sequence of samples. Fig. \ref{fig:ag_trellis} illustrates the trellis diagram of the Viterbi algorithm. For each transition, the computation of the glitch error metric requires $L+1$ multiplications and $L-1$ additions. The number of states at time index $m$ is the number of representations for $x[m]$, i.e., $|\mathcal{R}\big(x[m]\big)|$. Therefore, the computational complexity of the dynamic programming approach is $\mathcal{O}(S^2 L)$ per sample where $S$ is the average number of representations per input. Let's now assume that each binary representation of dimension $L$ gives us a number between 0 and $2^N-1$. Consequently, $S = 2^{L-N}$ and the complexity is $\mathcal{O}(4^{L-N} L)$; thus, the complexity of the proposed algorithm increases exponentially with $L-N$. Therefore, a solution with a reasonable complexity is needed to take advantage of the optimized architecture.

\textit{2) Best next greedy mapping}: A greedy algorithm that selects the best representation at time index $m$ based on only the previous sample can be formulated as
\begin{equation}
\m{W}_\text{opt}(x[m]) = \argmin_{\m{W}\big(x[m]\big)} \Big(\myabs{\m{W}^\text{T}\big(x[m]\big) - \m{W}^\text{T}\big(x[m-1]\big)}\m{B}\Big)^2.
\end{equation}
Although the complexity is $\mathcal{O}(2^{L-N} L)$ per sample, the best representations can be calculated offline for each representation of the previous sample and the value of the current sample. From an implementation point of view, this approach can be implemented through \gls*{LUT} of size $2^{N+L}$. One drawback of this approach is that it still requires sequential processing of samples, i.e., parallelization is not possible.

\textit{3) Memoryless mapping}: A much less sophisticated algorithm is to only use one unique representation for each input codeword. Therefore, we need to find the best representations for all the inputs that minimize the total glitch error power. One can write the problem as
\begin{equation}
\argmin_{\m{W}(0),...,\m{W}(2^N-1)} \sum_{x=0}^{2^N-1} \sum_{y=0}^{2^N-1} \Prob{x,y} \Big(\myabs{\m{W}^\text{T}(y) - \m{W}^\text{T}(x)} \m{B}\Big)^2.
\end{equation}
The optimization problem can be solved iteratively. At each step, we fix the representations for all inputs and only find the best selection for $x$, i.e.,
\begin{equation}
\m{W}_\text{opt}(x) = \argmin_{\m{W}(x)} \sum_{y=0}^{2^N-1} \Prob{x,y} \Big( \myabs{\m{W}^\text{T}_\text{opt}(y) - \m{W}^\text{T}(x)} \m{B} \Big)^2.
\end{equation}
Note that once the optimization problem is solved, the hardware computational complexity is $\mathcal{O}(1)$ per sample, and samples can be processed independently. The memoryless mappings of a few input codes are presented in Table \ref{tab:te_single_rep}. As an example, a switch from code $127$ to $128$ requires one cell of weight $2$ to turn on and one cell of weight $1$ to turn off. For comparison, in a segmented DAC with $12$ switches, a transient from code $127$ to $128$ requires one cell of weight $32$ to turn on and five cells of weight $1$, $2$, $4$, $8$, and $16$ to turn off.

\begin{table}[!tb]
\vspace{2ex}
\footnotesize
\centering
\hspace{1ex}
\begin{tabular}{||l||*{12}{p{0.6em}}||}
\hline
  $\m{B}$ 	  &$1$&     $2$&     $4$&     $8$&    $11$&    $16$&    $20$&    $25$&    $27$&    $35$&    $48$&    $58$\\
\hline\hline
   $122$&     $0$&     $0$&     $1$&     $1$&     $0$&     $0$&     $0$&     $0$&     $1$&     $1$&     $1$&     $0$\\
   $123$&     $1$&     $0$&     $1$&     $1$&     $0$&     $0$&     $0$&     $0$&     $1$&     $1$&     $1$&     $0$\\
   $124$&     $0$&     $1$&     $1$&     $1$&     $0$&     $0$&     $0$&     $0$&     $1$&     $1$&     $1$&     $0$\\
   $125$&     $1$&     $1$&     $1$&     $1$&     $0$&     $0$&     $0$&     $0$&     $1$&     $1$&     $1$&     $0$\\
   $126$&     $0$&     $0$&     $0$&     $0$&     $0$&     $1$&     $0$&     $0$&     $1$&     $1$&     $1$&     $0$\\
   $127$&     $1$&     $0$&     $0$&     $0$&     $0$&     $1$&     $0$&     $0$&     $1$&     $1$&     $1$&     $0$\\
   $128$&     $0$&     $1$&     $0$&     $0$&     $0$&     $1$&     $0$&     $0$&     $1$&     $1$&     $1$&     $0$\\
   $129$&     $1$&     $1$&     $0$&     $0$&     $0$&     $1$&     $0$&     $0$&     $1$&     $1$&     $1$&     $0$\\
   $130$&     $0$&     $0$&     $1$&     $0$&     $0$&     $1$&     $0$&     $0$&     $1$&     $1$&     $1$&     $0$\\
   $131$&     $1$&     $0$&     $1$&     $0$&     $0$&     $1$&     $0$&     $0$&     $1$&     $1$&     $1$&     $0$\\
   $132$&     $0$&     $1$&     $1$&     $0$&     $0$&     $1$&     $0$&     $0$&     $1$&     $1$&     $1$&     $0$\\
   $133$&     $1$&     $1$&     $1$&     $0$&     $0$&     $1$&     $0$&     $0$&     $1$&     $1$&     $1$&     $0$\\
   $134$&     $0$&     $0$&     $0$&     $1$&     $0$&     $1$&     $0$&     $0$&     $1$&     $1$&     $1$&     $0$\\
\hline
\end{tabular}
\vspace{2ex}
\caption{Memoryless mapping of codewords $122$ to $134$ for the optimized basis of length $12$.}
\label{tab:te_single_rep}
\end{table}

\section{Simulation Results}

In this section, Matlab behavioral level simulation results are presented to evaluate the glitch performance of the proposed optimized architecture and to validate our theoretical analysis. An $8$-bit \gls*{DAC} is considered for all the investigations.

\begin{figure}[!tb]
\centering
\footnotesize
\psfrag{Hybrid, best_next}{Segmented DAC}
\psfrag{Optimized, single_rep_opt}{Optimized, memoryless}
\psfrag{Optimized, best_next}{Optimized, best next}
\psfrag{Optimized, viterbi}{Optimized, Viterbi}
\psfrag{Basis length}{Basis length $L$}
\psfrag{Normalized cost (Simulation)}{\hspace{5ex} Normalized metric}
\includegraphics[scale=\scalingFig]{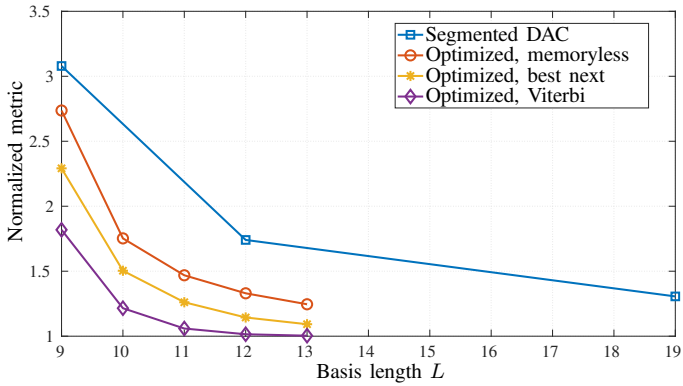}
\caption{Normalized glitch metric as a function of basis length $L$.}
\label{fig:ag_metric}
\end{figure}

Fig. \ref{fig:ag_metric} illustrates the glitch metric derived in Eq. (\ref{eq:ag_metric}) for a segmented \gls*{DAC} and the proposed optimized architecture. The $x$-axis is the number of basis elements and the $y$-axis represents the metric normalized by the metric of a thermometer-coded DAC. As evident in the figure, the proposed approach is advantageous over the same size segmented DAC. Three segmented DACs on the blue curve are $2$T+$6$B, $3$T+$5$B, and $4$T+$4$B architectures, where the first number represents the number of \gls*{MSB}s used for thermometer weighting and the second number denotes the number of binary bits. The $4$T+$4$B segmentation requires $19$ switches. However, the proposed architecture with 10 switches achieves better performance when used with the Viterbi algorithm. The greedy best next approach requires $11$ elements and the memoryless scheme needs $13$ elements to outperform the $4$T+$4$B segmented DAC.

\begin{figure}[!tb]
\centering
\footnotesize
\psfrag{Hybrid20, single_rep}{Segmented DAC}
\psfrag{Optimized20, single_rep_opt}{Optimized, memoryless}
\psfrag{Optimized20, best_next}{Optimized, best next}
\psfrag{Optimized20, viterbi}{Optimized, Viterbi}
\psfrag{basis_len}{Basis length $L$}
\psfrag{SDR [dB]}{SNDR [dB]}
\includegraphics[scale=\scalingFig]{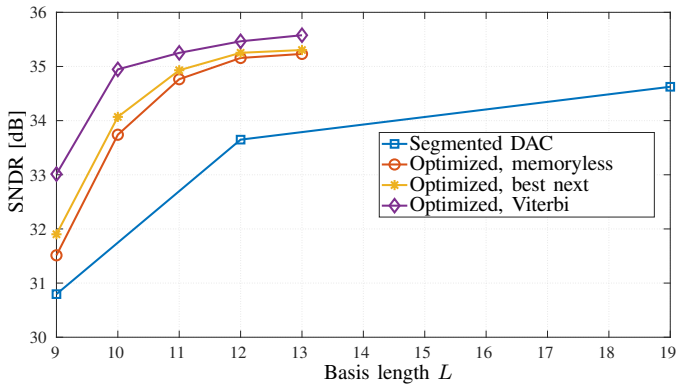}
\caption{SNDR as a function of basis length $L$.}
\label{fig:ag_sndr_vs_L}
\end{figure}

The \gls*{SNDR} achieved by each of the architectures is illustrated in Fig. \ref{fig:ag_sndr_vs_L}. Note that the \gls*{SNDR} does not include quantization noise and the only impairment simulated is the timing offset between on and off transients. The figure is consistent with the metric results in Fig. \ref{fig:ag_metric} and demonstrates the advantage of the optimized architectures over the segmented structure. 

\begin{figure}[!tb]
\centering
\footnotesize
\psfrag{8B}{$8$B}
\psfrag{4T+4B}{$4$T+$4$B}
\psfrag{Optimized L=13, single rep}{Opt. memoryless}
\psfrag{8T}{$8$T}
\psfrag{SFDR [dB]}{SFDR [dB]}
\psfrag{asymmetry_offset}{\hspace{-10ex} Normalized timing error $(\tau_\text{on/off}/T)$}
\includegraphics[scale=\scalingFig]{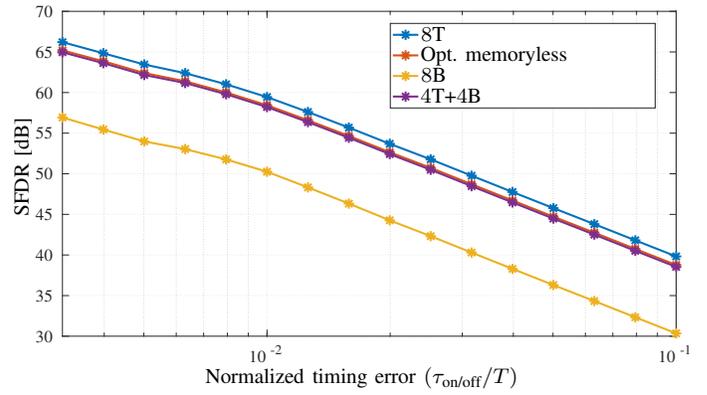}
\caption{SFDR comparison of the proposed architecture and segmentation.}
\label{fig:ag_sfdr}
\end{figure}

The \gls*{SFDR} for a sine waveform with a normalized frequency of $31/1024$ is illustrated in Fig. \ref{fig:ag_sfdr}. \gls*{SFDR} is a measure of DAC linearity that quantifies the power ratio between the fundamental tone and the most significant spurious tone at the output. The optimized approach with $13$ elements offers similar \gls*{SFDR} performance compared to the $4$T+$4$B segmented DAC. The binary and unary DACs are also plotted for comparison where they have the worst and best \gls*{SFDR} performances, respectively. 

\section{Conclusion}
DAC non-linear distortion is the leading degradation factor in high data rate applications such as telecommunication systems. In this paper, we focused on fall/rise asymmetry glitches and proposed a novel architecture that can outperform traditional segmented structures. We discussed how optimal mapping can be computed and proposed several greedy algorithms for efficient implementation. As the next step of this work, we will be performing transistor-level simulations to verify the potential benefit of the proposed architecture.

\bibliographystyle{IEEEtran}
\bibliography{IEEEabrv,main}

\end{document}